# A Constraints Fusion-induced Symmetric Nonnegative Matrix Factorization Approach for Community Detection

Zhigang Liu[1,2], Xin Luo[3*]

[1]*College of Computer Science and Technology, Chongqing University of Posts and Telecommunications, Chongqing, 400065, China*
[2]*Chongqing Key Laboratory of Big Data and Intelligent Computing, Chongqing Institute of Green and Intelligent Technology, Chinese Academy of Sciences, Chongqing, 400714, China*
[3]*College of Computer and Information Science, Southwest University, Chongqing 400715, China*

**Abstract**

Community is a fundamental and critical characteristic of an undirected social network, making community detection be a vital yet thorny issue in network representation learning. A symmetric and non-negative matrix factorization (SNMF) model is frequently adopted to address this issue owing to its great interpretability and scalability. However, it adopts a single latent factor matrix to represent an undirected network for precisely representing its symmetry, which leads to loss of representation learning ability due to the reduced latent space. Motivated by this discovery, this paper proposes a novel Constraints Fusion-induced Symmetric Nonnegative Matrix Factorization (CFS) model that adopts three-fold ideas: a) Representing a target undirected network with multiple latent factor matrices, thus preserving its representation learning capacity; b) Incorporating a symmetry-regularizer that preserves the symmetry of the learnt low-rank approximation to the adjacency matrix into the loss function, thus making the resultant detector well-aware of the target network's symmetry; and c) Introducing a graph-regularizer that preserves local invariance of the network's intrinsic geometry, thus making the achieved detector well-aware of community structure within the target network. Extensively empirical studies on eight real-world social networks from industrial applications demonstrate that the proposed CFS model significantly outperforms state-of-the-art models in achieving highly-accurate community detection results.

## 1. Introduction

Numerous entities are ubiquitous in real-world applications, which along with their interactions can be represented as a large-scale undirected network, such as a rating network in recommended systems [1-3], a biological network in bioinformatics [4-6], and a social network in social computing [7-10]. One of the most prominent features of a network is underlying community structure, which plays a critical role in extraction of useful knowledge that describes its networking organization and functional mechanism [11,12]. In general, a community is taken as a special sub-graph that contains a group of nodes with tight connections to form a compact cluster, while the inter-community connections are relatively sparse [13-15]. Aiming at revealing the latent structure indicating groups of cohesive nodes in a network, community detection has become a challenging problem in network representation learning and drawn much interest from different disciplines.

Commonly, community detection can be considered as a problem of graph clustering, which is related to various real applications, such as outlier detection [16], social recommendation [17], and biological module identification [18]. To address it, a pyramid of detection approaches have been proposed [19,20]. Generally, the existing community detection methods are categorized as heuristic-based and model-based ones [21]. By meeting some predefined evaluation metrics, e.g., modularity [22], a heuristic method is able to identify communities via searching groups of nodes in a target network. Note that modularity is one of the most widely-accepted metrics for community detection indicating the strength of division of a network into clusters, i.e., it measures the difference between the number of edges falling within clusters in the target network and the corresponding expected number in an equivalent network where edges are distributed randomly. Thus, many methods have been proposed to discover communities via maximizing modularity, e.g., Fast Newman algorithm [23] and fast unfolding [24]. In addition, many studies leverage particular substructures, e.g., clique, as a heuristic measurement to perform community detection [25].

More recent methods for community detection are model-based ones that consider identifying communities as a learning task, i.e., network representation learning. To date, many machine learning techniques have been successfully applied to community detection, e.g., spectral analysis [26,27], matrix factorization [8,21], and deep learning [9,20,28]. Among those sophisticated methods, a Nonnegative Matrix Factorization (NMF)-based model is one of the representative approaches for community detection on undirected networks from real world applications. Given an undirected network, a standard NMF-based community detection method aims to learn a low-rank representation for its adjacency matrix by factorizing it into two latent factor matrices that are called the basis matrix and the representation matrix, respectively. NMF is quite suitable for graph clustering tasks. The reasons are two-fold: a) its potential clustering capability, i.e., NMF and its variants have been proven to be equivalent to some classical clustering methods such as $k$-means; and b) its nice interpretability to community structure. Most of the current NMF-based community detectors achieve better performance via improving an NMF model's representation learning ability. For instance, Wang et al. [29] propose to apply the NMF-based approaches for community detection on three types of networks including the undirected network, directed network, and compound network. Sun et al. [30] propose a nonnegative encoder-decoder-based model for community detection, which introduces an encoder to transform the adjacency matrix into a low-dimensional representation, and a decoder to reconstruct the network from the learnt representation, thereby improving its representation ability. Ma et al. [31] propose to incorporate the prior information into NMF's factorization process to improve its performance for community detection on a multi-layer network. Despite of their efficiency in community detection, these methods fail to correctly describe the symmetry of an undirected network, which is its inherently fundamental property.

To model the symmetry of an undirected network, a Symmetric and Nonnegative Matrix Factorization (SNMF) model is developed, which works by learning a unique latent representation matrix $X$ to construct a low-rank approximation $XX^T$ to the adjacency matrix of the target network, thereby correctly characterizing its symmetry. The previous work [32] is proven that it is equivalent to a spectral clustering method. Yang et al. [33] first present a unified interpretation to an SNMF model in community detection, and then propose a unified semi-supervised framework that combines both network topology and prior information to improve detection accuracy. Shi et al. [34] propose a pairwisely constrained SNMF model that incorporates the pairwise constraints generated from a bit of ground-truth information to enhance its performance. In order to extract the intrinsic constraint information from the connections itself, Ye et al. [35] propose a homophily-preserving NMF model that combines both the network topology and node homophily of a network, thus achieving better representation of community structures. Luo et al. [36] propose several linear and nonlinear control schemes to adjust scaling-factors of nonnegative multiplicative update (NMU) learning rules, resulting in several highly-accurate SNMF-based community detectors. The above-mentioned SNMF-based methods achieve state-of-the-art performance. However, they only adopt a unique latent factor matrix to represent a target network, which ensures its rigorous symmetry but restricts its representation learning ability due to the shrinking of the latent space.

To tackle this issue, some efforts have been done to enhance an SNMF model's representation learning capacity. Kuang et al. [37] propose a nonsymmetric formulation for an SNMF model (called SymNMF) by introduce a constraint term that minimizes the differences between two latent factor matrices of an NMF model for the graph clustering task. By doing so, the proposed SymNMF model is optimized via a two-block coordinate descent algorithm, since it forces the separation of unknown variables by associating the two latent factors with two different matrices. Subsequently, Zhu et al. [38] and Li et al. [39] also transfer the SNMF to a penalized nonsymmetric NMF model by introducing an equality regularizer to force its two latent factors identical and balancing the effect of the regularizer and the error term with a tunable hyper-parameter. They present an alternating nonnegative least squares (ANLS)-type learning algorithm to solve the reformulation model, as well as the rigorously mathematical proof of its convergence. Although the above methods can represent partial symmetry via setting a large enough regularization coefficient to force the two latent factors identical, the balance between symmetry and representation learning capacity is difficult to strike perfectly. With a small regularization coefficient, the models cannot represent the symmetry effectively. However, with a large regularization coefficient, the representation of symmetry is naturally enhanced, but the optimization process would overly concern the regularizer while weakening the fitting of the error term, which impacts the accuracy of the obtained network representation. Besides, if the solutions of the two latent factor matrices are close enough, the models are actually degenerated to SNMF. Motivated by this discovery, this paper presents and aim to answer the following question:

**Is it possible to develop an NMF-based community detector that is able to represent a target undirected network's symmetry while preserving its representation learning capacity, thus obtaining highly accurate community detection?**

To answer this question, in this paper, we propose a <u>C</u>onstraints <u>F</u>usion-induced <u>S</u>ymmetric Nonnegative Matrix Factorization (CFS) model that adopts the following three ideas: a) Representing a target undirected network with multiple latent factor matrices, thus preserving its representation learning capacity; b) Incorporating a symmetry-regularizer that preserves the symmetry of the learnt low-rank approximation to the adjacency matrix into the loss function, thus making the resultant detector well-aware of the target network's symmetry; and c) Introducing a graph-regularizer that preserves local invariance of the network's intrinsic geometry, thus making the achieved detector well-aware of community structure within the target network.

Main contributions of this work include:
a) **A novel symmetry-regularizer.** It realizes the symmetry-constraint by forcing the resultant low-rank approximation and its transpose matrix identical. Note that it does not force the identity of two involved latent factor matrices directly. Hence, it will not impact the model's representation learning capacity.
b) **A CFS model.** It models a target undirected network's symmetry by incorporating the proposed symmetry-regularizer which can preserve its representation learning capacity, thus achieving good enhancement of the representation learning ability for an SNMF model. In addition, it further incorporates the graph-regularization into its learning objective to preserve the intrinsic geometric features of the target network, thus ensuring its learning ability regarding local invariance.
c) **Rigorous convergence proof of CFS.** The proof consists of two separate steps: a) Proving the non-increasing tendency of the learning objective by leveraging an auxiliary function-incorporated method; and b) Proving the latent factor learning sequences' convergence to a stationary point of the learning objective by proving the establishment of its KKT conditions.

Experimental results on eight publicly available and real networks form real applications indicate that the proposed CFS-based community detector obtains significant accuracy gain comparing with the benchmark and state-of-the-art methods.

## 2. Preliminaries

*2.1. Problem Formulation*

A community detector models a target network with a graph $G=(V, E)$ where $V=\{v_i \mid i=1, …, n\}$ is a set of $n$ nodes and $E=\{e_{ij} \mid i, j=1, …, n\}$ is a set of $m$ edges. Then $G$ is denoted by its adjacency matrix $A=[a_{ij}]\in\mathbb{R}^{n\times n}$, where $a_{ij}$ characterizes the interaction relationship between nodes $v_i$ and $v_j$. In detail, for an unweighted network, $a_{ij}$ is one if $e_{ij}\in E$, and zero otherwise. For a weighted network, $A$ is real-valued. Besides, $A$ is symmetric if $G$ is an undirected graph, and asymmetric otherwise. In this study, we concern the problem of community detection in undirected networks, where all relationships between pairwise entities are symmetric. Given a network $G=(V, E)$, $\forall v_i\in V$, generally, a community detector aims to identify its correct affiliation to a community that consists of its closely related nodes in the target network.

Given a network $G=(V, E)$, an NMF-based community detector generally works by the following four key process stages: a) constructing the adjacency matrix $A$ according to $G$; b) building an NMF-based model to learn a low-rank approximation to $A$; c) solving the model; and d) identifying node-community affiliation according to the obtained model. Specifically, it first assumes that $G$ has $K$ communities according to the prior information. Then, it learns a rank-$K$ approximation $\hat{A}$ to $A$ as $\hat{A}=UX^T$ subjecting to $U\geq 0$ and $X\geq 0$. From the perspective of linear expression, $U$ consists of a set of base vectors and $X$ is taken as a coefficient matrix. Hence, $X$ can be also considered as a representation matrix for the target network, where each row vector denotes a representation for a network node. In the context of NMF-based community detection, $X$ is taken as the soft threshold indicator to identify the node-community affiliation, i.e., $\forall j\in\{1, 2, …, n\}$ and $k\in\{1, 2, …, K\}$, each entry $x_{jk}$ indicates the probability of node $v_j$ belonging to community $C_k$, which can be formulated as [8, 36]:

$$\forall v_j \in V: v_j \in C_k, \text{ if } x_{jk} = \max\{x_{js} \mid s = 1, 2, …, K\}. \tag{1}$$

With (1), in terms of the case of non-overlapping community detection, for an arbitrary node $v_j$, we only need to assign it into the community $C_k$ by satisfying the condition that $x_{jk}$ is the maximum entry in row vector $x_j$.

*2.2. A (Symmetric)NMF-based Community Detector*

An NMF model has been widely used in information recommendation [40-46], dimensionality reduction [47-55], and feature extraction [55-60], owing to its nice interpretability. In addition, NMF is also highly suitable for the graph clustering task such as community detection, owing to its potential clustering capability. Given a target network $G=(V, E)$ with the adjacency matrix $A$, an NMF-based model aims to learn a rank-$K$ approximation $\hat{A}$ to $A$ with two different nonnegative latent factor matrices $U=[u_{ik}]\in\mathbb{R}^{n\times K}$ and $X=[x_{jk}]\in\mathbb{R}^{n\times K}$, i.e., yielding $\hat{A}=UX^T$. To acquire $U$ and $X$, we can solve an objective function that minimizes the difference between $A$ and $\hat{A}$. By adopting the square of the Frobenius norm of the difference between $A$ and $UX^T$, such an objective function is achieved:

$$\min \mathcal{J}(U, X) = \left\| A - UX^T \right\|_F^2, \text{ s.t. } U \geq 0, X \geq 0. \tag{2}$$

where $||\cdot||_F$ denotes the Frobenius norm. $U$ and $X$ are constrained to be nonnegative for the good interpretability that they together describe the nonnegative probabilities characterizing each node's community tendency. Note that $\mathcal{J}(U, X)$ is non-convex to $U$ and $X$ together, thus making their global optima intractable. Fortunately, their solution of stationary point can be obtained by leveraging an alternative and iterative algorithm [43,44,60-66]. For instance, [44] provides a nonnegative multiplicative update (NMU) algorithm to iteratively optimize (2) as:

$$u_{ik} \leftarrow u_{ik} (AX)_{ik} / (UX^TX)_{ik}, \quad x_{jk} \leftarrow x_{jk} (A^TU)_{jk} / (XU^TU)_{jk}. \tag{3}$$

With (3), a benchmark NMF-based community detector is established [44]. However, although such a model learns a network's representation via adopting multiple latent factors, it does not consider the symmetry of the target undirected network, which is its intrinsic property.

For an undirected network $G$, its adjacency matrix $A$ is a symmetric square matrix. To better represent its symmetry, an NMF model can be transformed to the symmetric NMF (SNMF) with a symmetric factorization form, i.e., $\hat{A}=UU^T$, by making two latent factor identical. In other words, an SNMF model adopts a unique latent factor matrix for representing the target network, thereby guaranteeing its rigorous symmetry. Similarly, its objective function is given as:

$$\min \mathcal{J}(U) = \left\| A - UU^T \right\|_F^2, \text{ s.t. } U \geq 0, \tag{4}$$

where $U$ denotes the community soft threshold indicator, and it can be obtained by an NMU learning algorithm, i.e.,

$$u_{ik} \leftarrow u_{ik} (AU)_{ik} / (UU^TU)_{ik}. \tag{5}$$

As revealed in [47], the convergence of an SNMF model under the learning scheme in (5) can be commonly unstable, making an SNMF-based community detector suffer from accuracy reduction. Following the suggestion in [47], we adopt a learning scheme with a linearly adjusted multiplicative, i.e.,

$$u_{ik} \leftarrow u_{ik}\left(1/2+(AU)_{ik}/2(UU^{\mathrm{T}}U)_{ik}\right). \quad (6)$$

With (6), a benchmark SNMF-based community detector is established [29], which implies a strong symmetric constraint into its approximation, thereby ensuring the rigorous symmetry of the target network. Moreover, it has been proven to be equivalent to a spectral clustering model, making its nice interpretability for community detection which is actually a node clustering problem on graph data. Nevertheless, adopting a unique latent factor matrix gives rise to shrinkage of the latent space, which reduces its representation learning ability to a target network.

## 3. Methods

In this part, we aim to present our CFS-based community detector. Its main principle is three-fold:
a) Adopting an asymmetric factorization form with multiple latent factors to represent a target undirected network, thus preserving its representation learning capacity;
b) Leveraging a novel symmetry-regularizer that forces the symmetry of the resultant low-rank approximation to the adjacency matrix, thus preserving its representation learning ability for macroscopic symmetry of a target undirected network; and
c) Incorporating a graph-regularizer into its learning objective to preserve the intrinsic geometric features of the target network, thus ensuring its learning ability regarding local invariance.

The learning algorithm for solving the proposed model is achieved by following the principle of NMU. Rigorous mathematical proof is presented to indicate that the proposed algorithm enables a CFS model to converge to a KKT stationary point of its learning objective. In addition, detailed algorithm is designed for implementing the proposed CFS-based community detector, as well its complexity analysis is also provided.

### 3.1. A Symmetry-regularized NMF Model

To well represent the symmetry of a target network, a novel symmetry-regularizer, i.e., $||UX^{\mathrm{T}}-(UX^{\mathrm{T}})^{\mathrm{T}}||_F^2$, is introduced to penalize the symmetry of the learnt low-rank approximation. By doing so, the learning objective function of CFS is given as:

$$\min \mathcal{J}(U,X) = \left\|A - UX^{\mathrm{T}}\right\|_F^2 + \frac{\mu}{2}\left\|UX^{\mathrm{T}} - XU^{\mathrm{T}}\right\|_F^2, \quad \text{s.t.} \ U \geq 0, X \geq 0, \quad (7)$$

where $A$ is the adjacency matrix of the target network. $U$ and $X$ are desired latent factor matrices, and they together form the low-rank approximation to $A$, i.e., $UX^{\mathrm{T}}$. $\mu>0$ is a scale parameter being taken as a tradeoff between the generalized error and the symmetry regularization. With the adopted symmetry-regularizer in (7), the resultant low-rank approximation and its transpose are forced to be identical, which brings about the target network's symmetry. Besides, such a symmetry constraint is loose and can be adjusted with varying $\mu$. Noted that by adopting the asymmetric factorization form in (7) to construct an SNMF model, its latent space is not shrunk with two different latent factor matrices. Hence, (7) also enables a CFS model to obtain high representation learning capacity to $A$.

To preserve the local invariance of the target network, we further incorporate the graph regularization into (7) to achieve:

$$\min \mathcal{J}(U,X) = \left\|A - UX^{\mathrm{T}}\right\|_F^2 + \frac{\mu}{2}\left\|UX^{\mathrm{T}} - XU^{\mathrm{T}}\right\|_F^2 + \lambda \mathrm{tr}(X^{\mathrm{T}}LX), \quad \text{s.t.} \ U \geq 0, X \geq 0, \quad (8)$$

where tr(·) denotes the trace of matrix. $\lambda>0$ is another tunable parameter to adjust the effect of graph-regularization. $L=D-W$ is the Laplacian matrix of $A$, where $W$ is a similarity matrix measuring the affinity between each node pair, and $D$ is the degree matrix computed by $D_{ii}=\sum_l W_{il}$. In this work, we make $W$ and $A$ numerically equal according to [48]. Based on the commonly accepted property of $||X||_F^2 = \mathrm{tr}(XX^{\mathrm{T}})$, we achieve the following deduction:

$$\min \mathcal{J}(U,X) = \mathrm{tr}(AA^{\mathrm{T}} - 2AXU^{\mathrm{T}} + UX^{\mathrm{T}}XU^{\mathrm{T}}) + \mu\mathrm{tr}(UX^{\mathrm{T}}XU^{\mathrm{T}} - UX^{\mathrm{T}}UX^{\mathrm{T}}) + \lambda\mathrm{tr}(X^{\mathrm{T}}LX), \quad \text{s.t.} \ U \geq 0, X \geq 0, \quad (9)$$

where the principle of tr($AB$)=tr($BA$) is adopted in the second step of (9). By letting $\Phi=[\phi_{ik}]\in\mathbb{R}^{n\times K}$ and $K=[\kappa_{jk}]\in\mathbb{R}^{n\times K}$ be the Lagrangian multiplier matrices for the nonnegative constraints: $U=[u_{ik}]\geq 0$ and $X=[x_{jk}]\geq 0$, we construct the Lagrangian function as:

$$\mathcal{L}(U,X) = \mathrm{tr}(AA^{\mathrm{T}} - 2AXU^{\mathrm{T}} + UX^{\mathrm{T}}XU^{\mathrm{T}}) + \mu\mathrm{tr}(UX^{\mathrm{T}}XU^{\mathrm{T}} - UX^{\mathrm{T}}UX^{\mathrm{T}}) + \lambda\mathrm{tr}(X^{\mathrm{T}}LX) - \mathrm{tr}(\Phi U^{\mathrm{T}}) - \mathrm{tr}(KX^{\mathrm{T}}), \quad (10)$$

whose partial derivatives with respect to $U$ and $X$ are computed as:

$$\frac{\partial \mathcal{L}(U,X)}{\partial U} = -2AX + 2(1+\mu)UX^{\mathrm{T}}X - 2\mu XU^{\mathrm{T}}X - \Phi, \quad (11a)$$

$$\frac{\partial \mathcal{L}(U,X)}{\partial X} = -2A^{\mathrm{T}}U + 2(1+\mu)XU^{\mathrm{T}}U - 2\mu UX^{\mathrm{T}}U + 2\lambda LX - K. \quad (11b)$$

Hence, a local minimum of (10) is achieved by setting (11) and (12) to be zeros, i.e., $\partial\mathcal{L}(U,X)/\partial U=0$ and $\partial\mathcal{L}(U,X)/\partial X=0$. After that, based on the KKT conditions that $\forall i,j\in\{1,2,\ldots,n\}, k\in\{1,2,\ldots,K\}$: $\phi_{ik}u_{ik}=0$ and $\kappa_{jk}x_{jk}=0$, the optimal solutions of $u_{ik}$ and $x_{jk}$ are obtained:

$$-\left(AX+\mu XU^{\mathrm{T}}X\right)_{ik}u_{ik}+(1+\mu)\left(UX^{\mathrm{T}}X\right)_{ik}u_{ik}=0 \Rightarrow u_{ik}=u_{ik}\frac{\left(AX+\mu XU^{\mathrm{T}}X\right)_{ik}}{(1+\mu)\left(UX^{\mathrm{T}}X\right)_{ik}}; \quad (12a)$$

$$-\left(A^{\mathrm{T}}U+\mu UX^{\mathrm{T}}U+\lambda AX\right)_{jk}x_{jk}+(1+\mu)\left(XU^{\mathrm{T}}U\right)_{jk}+\lambda\left(DX\right)_{jk}=0 \Rightarrow x_{jk}=x_{jk}\frac{\left(A^{\mathrm{T}}U+\mu UX^{\mathrm{T}}U+\lambda AX\right)_{jk}}{(1+\mu)\left(XU^{\mathrm{T}}U\right)_{jk}+\lambda\left(DX\right)_{jk}}. \quad (12b)$$

Finally, with (13) and (14), the following NMU-type learning scheme for CFS is achieved:

$$u_{ik}^{(t+1)} \leftarrow u_{ik}^{(t)}\left(AX+\mu XU^{\mathrm{T}}X\right)_{ik}\Big/(1+\mu)\left(UX^{\mathrm{T}}X\right)_{ik}, \quad (13a)$$

$$x_{jk}^{(t+1)} = x_{jk}^{(t)}\left(A^{\mathrm{T}}U+\mu UX^{\mathrm{T}}U+\lambda AX\right)_{jk}\Big/\left((1+\mu)\left(XU^{\mathrm{T}}U\right)+\lambda DX\right)_{jk}. \quad (13b)$$

Hereto, with (13) a CFS-based community detector is established. It should be pointed that if $\mu$ is set to zero, (13) degrades into the learning scheme for a graph-regularized NMF (GNMF)-based community detector [33]. If both $\mu$ and $\lambda$ are set to zeros, (13) degrades into the learning scheme for a benchmark NMF-based community detector [44].

## 4. Experimental Results and Analysis

### 4.1 General Settings

*Evaluation Protocol.* To obtain fair experimental results, we adopt two types of metrics for evaluating the detection accuracy of tested community detectors. Note that as an unsupervised learning process, the performance of a community detection method commonly tested via leveraging the ground-truth, i.e., node-community labels, to measure its detecting accuracy. Hence, to avoid substituting label information into the model building process, we adopt an evaluation metric without ground-truth as a validation means to tune hyper-parameters. By doing so, a final community detector with the optimum hyper-parameters is established. Then, we test its performance by using several ground-truth based evaluation metrics. Hence, we first adopt an evaluation metric without ground-truth, i.e., **Modularity** [50], as a performance indicator to select hyper-parameters. We then adopt two ground-truth based evaluation metrics, i.e., Normalized Mutual Information (NMI) [51] and Adjusted Rand Index (ARI) [51] to test the performance of all involved community detectors.

*Datasets.* Eight networks collected form real-world applications are used to evaluate the performance of involved methods. As summarized in Table 1, all adopted datasets are popular, publicly-available and wide-used in prior and related studies.

Table 1 Details of Experimental Datasets

| No. | Datasets | #Nodes | #Edges | #Communities | Description |
|---|---|---|---|---|---|
| D1 | Amazon | 5,304 | 16,701 | 85 | Amazon product [73] |
| D2 | DBLP | 12,547 | 35,250 | 4 | DBLP collaboration [73] |
| D3 | Friendster | 11,023 | 280,755 | 13 | Friendster online [73] |
| D4 | LJ | 7,181 | 253,820 | 30 | LiveJournal online [73] |
| D5 | Orkut | 11,751 | 270,667 | 5 | Orkut online [73] |
| D6 | Youtube | 11,144 | 36,186 | 40 | Youtube online [73] |
| D7 | Polblogs | 1,490 | 16,718 | 2 | Blogs about US politics [72] |
| D8 | Cora | 2,708 | 5,429 | 7 | LINQS [74] |

*Models.* We compare the proposed CFS (M10) model with nine models including two benchmark models, i.e., NMF (M1) [44] and SNMF (M2) [47], and seven state-of-the-art models, i.e., NSED (M3) [30], SymNMF (M4) [37], GNMF (M5) [33], GSNMF (M6) [33], HPNMF (M7) [35], LpNMF (M8) [53] and SymHALS (M9) [38,39].

### 4.2 Sensitivity Analysis

Since hyper-parameters are commonly significant for a machine learning model [67-72], we aim to make clear the effects of the involved hyper-parameters in CFS, i.e., $\lambda$ and $\mu$. It should be pointed that their self-adaptation is still a critical and challenging issue, and we plan to address this issue in our future work. Note that in this part we leverage a ground-truth independent evaluation metric, i.e., Modularity, to tune hyper-parameters to avoid the issue of leaky validation.

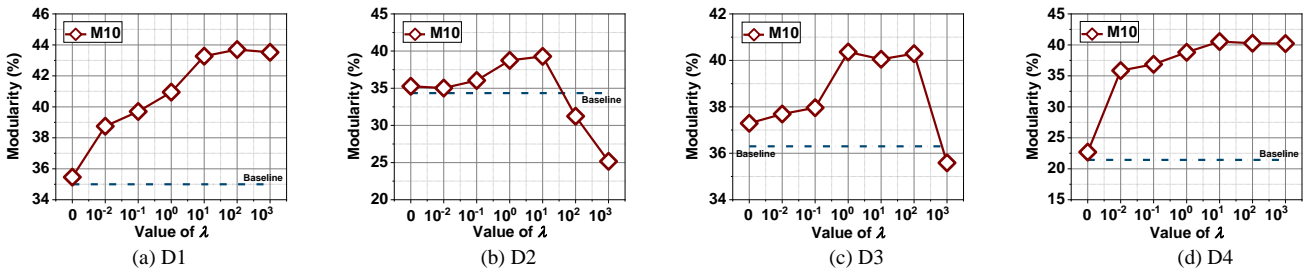

**Fig. 1.** The modularity values about various $\lambda$ on D1-D4.

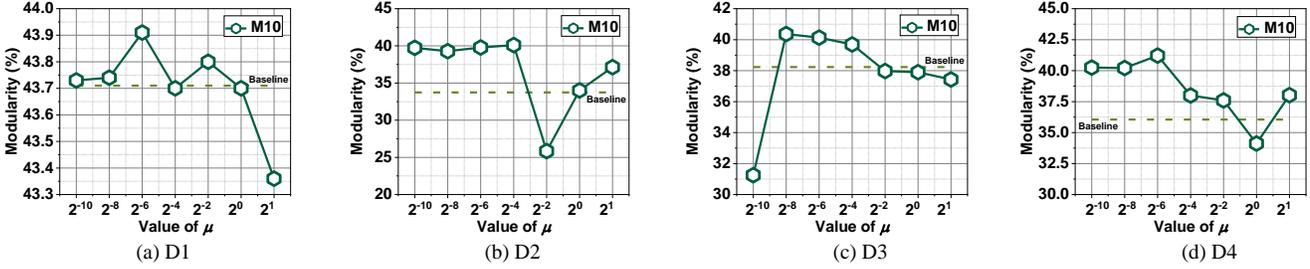

**Fig. 2.** The modularity values about various $\mu$ on D1-D4.

We first evaluate $\lambda$'s effects by tuning it in the set of $\{0, 10^{-2}, 10^{-1}, 10^0, 10^1, 10^2, 10^3\}$ while fixing $\mu$ at $2^{-8}$ randomly, as NMF is the baseline. The validation results regarding $\lambda$ on D1-D4 is depicted in Fig. 1. Then, we evaluate the effects of $\mu$ by tuning it in the set of $\{2^{-10}, 2^{-8}, 2^{-6}, 2^{-4}, 2^{-2}, 2^0, 2^1\}$ while setting the optimal value of $\lambda$ on each dataset, as GNMF is the baseline. Fig. 2 shows the validation results regarding $\mu$ on D1-D4. Form the experimental results shown in Fig. 1, Fig. 2, and Table 3, we see that the performance of M10 measured by Modularity is sensitive to both $\lambda$ and $\mu$. Both $\lambda$ and $\mu$ have vital effects on M10's detection accuracy. Their optimal values are datasets-dependent. However, when $\lambda=10$, $\mu\in[2^{-6}, 2^{-4}]$, M10 obtains the relatively better results.

*4.3 Performance Comparison*

We aim to compare our CFS with the baseline and state-of-the-art models, whose details are summarized in Section IV-A, to validate its performance. We record the average NMI and ARI values of ten compared models on eight real social networks in Tables 2 and 3. From the results, we conclude that:

a) **The proposed symmetry-regularizer is effective.** As shown in Table 2, we can see that M10 outperforms M5 on all eight testing cases in terms of NMI. According to the results in Table 3, M10 also outperforms M5 on seven testing cases out of eight in total in terms of ARI. Hence, the results discussed above tell us that the proposed symmetry-regularizer enable M10 to achieve more accurate detection accuracy. That is to say, the proposed symmetry-regularizer is effective.

b) **M10 outperforms its peers in achieving highly accurate community detection results.** From the comparison results recorded in Tables 2 and 3, we observe that M10 achieves the best detection accuracy on most testing cases across different evaluation metrics. Taking NMI as an example, we see that M10 significantly outperforms all the baseline and state-of-the-art models on six testing cases out of eight in total as shown in Table 2. Its accuracy gain is evident. Besides, the average Friedman rank of M10 regarding NMI is 1.25, which is the lowest one among all compared models as shown in Table 4, indicating that M10 achieves the best performance among all models. Overall, the comparison results indicate the significant superiority in detection accuracy of the proposed C-SNMF-based community detector compared both the baseline and state-of-the-art methods, owing to its compatible constraints fusion scheme via incorporating both the symmetry-regularizer and graph-regularizer.

Table 2. Community Detection Performance (NMI%±STD%) of Compared Methods on Each Dataset. (o Indicates That M10 Loses to the Comparison Models)

| NMI | M1 | M2 | M3 | M4 | M5 | M6 | M7 | M8 | M9 | M10 |
|---|---|---|---|---|---|---|---|---|---|---|
| D1 | 43.11±0.93 | 46.25±0.58 | 42.78±1.37 | 48.23±0.93 | 68.20±1.98 | 62.83±2.25 | **77.03±0.71**o | 68.20±1.48 | 53.59±1.02 | 70.02±3.33 |
| D2 | 6.51±1.81 | 7.19±2.05 | 5.97±2.52 | 9.30±1.13 | 4.98±2.31 | 10.42±3.04 | 15.07±5.79 | 13.34±3.44 | 11.97±4.29 | **20.92±3.72** |
| D3 | 62.48±10.52 | 72.18±5.26 | 69.23±3.02 | 80.59±0.00 | 73.38±6.24 | 76.36±4.60 | 77.01±4.53 | 86.42±7.58 | 63.63±6.83 | **88.26±3.56** |
| D4 | 24.30±1.71 | 33.81±2.37 | 19.38±4.60 | 41.74±0.79 | 49.64±3.45 | 72.34±5.43 | 63.96±1.39 | 51.12±5.51 | 42.01±2.74 | **73.13±1.82** |
| D5 | 29.76±0.60 | 28.17±1.32 | 33.53±6.11 | 32.12±6.38 | 31.89±6.28 | 71.16±11.58 | 46.70±11.27 | 44.81±13.37 | 39.06±4.88 | **71.79±8.32** |
| D6 | 17.41±1.88 | 17.53±0.57 | 17.62±1.27 | 19.38±1.40 | 18.80±1.02 | 42.75±2.66 | 30.40±2.53 | 22.70±5.27 | 11.49±1.03 | **50.83±3.44** |
| D7 | 48.02±3.76 | 45.10±1.40 | 47.46±3.35 | 45.40±1.42 | 48.21±3.74 | 45.10±1.40 | 44.71±1.21 | 44.98±0.00 | 47.05±1.11 | **50.38±1.46** |
| D8 | 8.55±4.25 | 19.21±2.57 | 11.18±5.31 | 16.08±3.89 | 11.89±2.82 | 17.78±1.29 | 18.67±2.12 | **21.14±0.00**o | 17.65±1.12 | 20.20±3.83 |
| Win/Loss | 8/0 | 8/0 | 8/0 | 8/0 | 8/0 | 8/0 | 7/1 | **7/1** | 8/0 | -- |
| Ranks* | 8.38 | 7.31 | 7.88 | 6.00 | 6.06 | 4.19 | 3.75 | 3.81 | 6.38 | **1.25** |
| *p*-value** | **0.0039** | **0.0039** | **0.0039** | **0.0039** | **0.0039** | **0.0039** | **0.0273** | **0.0078** | **0.0039** | -- |

*A lower Friedman rank value indicates a higher community detection accuracy.
**The accepted hypotheses with a significance level of 0.05 are highlighted.

Table 3. Community Detection Performance (ARI%±STD%) of Compared Methods on Each Dataset. (o Indicates That M10 Loses to the Comparison Models)

| ARI | M1 | M2 | M3 | M4 | M5 | M6 | M7 | M8 | M9 | M10 |
|---|---|---|---|---|---|---|---|---|---|---|
| D1 | 49.75±2.74 | 57.54±0.75 | 46.57±0.82 | 57.79±0.92 | 59.72±1.59 | 54.24±2.76 | **68.26±1.89**o | 61.75±1.56o | 60.87±2.49o | 60.84±1.88 |
| D2 | 19.38±3.46 | 19.35±4.34 | 17.08±4.76 | 22.44±5.12 | 18.20±3.95 | 24.37±6.73 | 28.16±7.37 | 24.22±4.27 | 25.58±8.79 | **38.33±3.22** |
| D3 | 63.53±7.76 | 70.96±6.41 | 69.46±2.14 | 78.54±0.00 | 74.06±3.74 | 69.28±1.23 | 70.36±4.24 | 80.45±17.56 | 54.45±9.07 | **85.78±5.56** |
| D4 | 46.64±5.14 | 59.41±6.94 | 43.65±5.30 | 50.85±0.42 | 55.39±9.01 | 52.80±8.66 | 44.01±9.89 | 43.85±4.31 | 58.42±5.36 | **72.42±5.17** |
| D5 | 33.08±1.94 | 34.77±2.22 | 38.00±6.52 | 36.78±7.12 | 36.59±6.88 | 65.99±12.05 | 53.46±10.64 | 50.63±13.78 | 40.37±8.94 | **74.76±5.93** |
| D6 | 28.83±1.72 | 28.26±0.58 | 28.82±2.01 | 30.06±0.94o | 29.65±1.02o | 26.08±2.11 | **31.57±7.19**o | 6.98±2.05 | 24.16±1.82 | 29.48±1.58 |
| D7 | 53.08±7.59 | 54.56±1.43 | 52.49±7.38 | 53.66±1.95 | 53.42±7.40 | 54.56±1.43 | 53.65±1.12 | 46.83±0.00 | 57.49±1.24 | **60.72±1.31** |
| D8 | 17.34±2.83 | 22.89±4.23 | 16.17±3.16 | 19.89±3.04 | 16.05±2.26 | 23.44±1.61 | 23.83±1.98o | **24.15±0.00**o | 22.09±2.26 | 23.60±1.58 |
| Win/Loss | 8/0 | 8/0 | 8/0 | 7/1 | 7/1 | 8/0 | 5/3 | 6/2 | 7/1 | -- |
| Ranks | 7.88 | 5.81 | 8.38 | 5.25 | 6.25 | 5.31 | 3.63 | 5.38 | 5.13 | **2.00** |
| *p*-value | **0.0039** | **0.0039** | **0.0039** | **0.0078** | **0.0078** | **0.0039** | 0.0742 | **0.0195** | **0.0078** | -- |

## 5. Conclusions

In the context of community detection in social networks, the symmetry of an undirected network plays an important role in correctly describing its intrinsically structural features. An SNMF algorithm introduces the symmetry constraint into an NMF framework, and it strictly models the symmetry of a given symmetric matrix. However, by doing so, its LF space reduces into half of the original NMF model, which degrades its performance for community detection. To address this issue, this work presents a novel Constraints Fusion-induced Symmetric Nonnegative Matrix Factorization (CFS) model. It realizes the relaxed symmetry by introducing an equality-constraint on the leant low-rank approximation of the target social network. Besides, it leverages a graph-regularization method into CFS to extract the features concerning the intrinsically geometric structure of networks. Empirical studies on eight real-world social networks from industrial applications demonstrate that a CFS-based detector can achieve the higher accuracy for community detection than state-of-the-art models. As future work, we plan to address the issue of self-adaptive hyper-parameters, self-adaption schemes, e.g., particle swarm optimization (PSO) algorithms [45,56,62] are highly demanded.